\documentclass{article} 
\usepackage{iclr2020_conference,times}

\usepackage{amsmath,amsfonts,bm}

\def\eqref#1{equation~\ref{#1}}

\def\1{\bm{1}}

\DeclareMathAlphabet{\mathsfit}{\encodingdefault}{\sfdefault}{m}{sl}
\SetMathAlphabet{\mathsfit}{bold}{\encodingdefault}{\sfdefault}{bx}{n}

\usepackage{hyperref}
\usepackage{url}

\title{Diagnosis of Diabetic Retinopathy in Ethiopia: Before the Deep Learning based Automation}

\author{Misgina Tsighe Hagos \\
Artificial Intelligence Directorate\\
Ethiopian Biotechnology Institute\\
Addis Ababa, Ethiopia \\
\texttt{tsighemisgina@gmail.com}
}

\iclrfinalcopy 
\begin{document}

\maketitle

\begin{abstract}

Introducing automated Diabetic Retinopathy (DR) diagnosis into Ethiopia is still a challenging task, despite recent reports that present trained Deep Learning (DL) based DR classifiers surpassing manual graders. This is mainly because of the expensive cost of conventional retinal imaging devices used in DL based classifiers. Current approaches that provide mobile based binary classification of DR, and the way towards a cheaper and offline multi-class classification of DR will be discussed in this paper.

\end{abstract}

\section{Introduction}
\label{introduction}

DR is a complication caused when diabetes damages blood vessels in the retina. It is one of the leading causes of vision loss in adults aged between 20 and 65 years. Globally, in 2010, 0.8 million out of 34 million blind people, and 3.7 million out of 191 million visually impaired people were caused by DR \citep{leasher2016global}. According to a 2015 report by IDF, there were more than 2.5 million cases of diabetes in Ethiopia \citep{idf2015}.

\subsection{Prevalence of Diabetic Retinopathy in Diabetic Patients}

To the best of the author’s knowledge, prevalence of DR in Ethiopia have not been studied before 2016. That started improving after the World Diabetes Foundation (WDF) in partnership with the Ethiopian Diabetes Association (EDA) underwent two projects on improving DR diagnosis process in Ethiopia, one of which equipped 12 hospitals, in different regions of the country, with resources for DR diagnosis \citep{wdf2019}. Various studies that targeted Ethiopian hospitals have shown high incidence of DR among diabetic people. By surveying studies performed in Oromia, Addis Ababa, SNNPR and Amhara regions, \citet{fite2019diabetic} showed that overall prevalence of DR among diabetes patients in Ethiopia was at 19.48\%.

\section{Current Approaches}
\label{approaches}

\subsection{Deep Learning based Classification}
\label{dl_based}

If DR is not treated at early stages, it can lead to permanent vision loss or impairment. In order to proceed with medical treatment, ophthalmologists employ different retinal imaging techniques such as fundus photography. DL have been employed for automated diagnosis of DR from retinal images, and it has proven successful \citep{gulshan2016development,abramoff2018pivotal,gulshan2019performance}. But these techniques have been seen to be expensive for diagnosis centers with limited budget\footnote{A leading diabetes and DR diagnosis center in Ethiopia, Diabetes Center at the Black Lion Hospital, isn’t currently providing DR diagnosis service because their fundus camera failed, and they couldn’t afford to replace or fix it.}. Shortage of trained ophthalmologists, and equipped healthcare centers are added challenges that affect availability of DR diagnosis in Ethiopia, especially the rural and remote areas.

\subsection{Mobile based Diagnosis}
\label{mobile_based}

A viable option to overcome the challenge with DL based diagnosis would be to incorporate a portable and cheaper retinal imaging device. A portable imaging device would make it easy to provide a point-of-care diagnosis. And, to circumnavigate shortage of professionals and diagnosis centers, DL algorithms could be used for automatic identification and classification of the disease.
Portable fundus cameras in combination with smartphones have been proposed and used for internet based \citep{rajalakshmi2018automated} and offline \citep{sosale2020medios} DR detection. While internet based diagnosis systems communicate with a central server to classify fundus images, offline diagnosis provides an instant service that is carried out on a smartphone, starting from image capturing until displaying diagnosis results. Offline methods are preferred to internet based DR diagnosis systems for areas with limited or no availability of internet connection. Standalone smartphone based offline DR diagnosis methods that provide binary classification have been developed in \citet{hagos2019automated} and \citet{sosale2020medios}. Smartphone based DR diagnosis systems require a handheld retinal image capturing device to be installed on a host phone's camera. The Remidio Fundus on Phone (FOP) Non-mydriatic device, which was used for experimentation in \citet{sosale2020medios}, and oDocs nun \citep{oDocsnun2018} are some of the smartphone based fundus cameras on market.

Even though portable fundus cameras can be cheaper than conventional fundus cameras, one downside they have is the reduced quality of retinal images they capture \citep{gosheva2017quality}. With a reduced image quality, let alone employing DL for identification, the chances of an ophthalmologist manually diagnosing for DR could get lowered \citep{yao2016generic}. If image quality of portable fundus cameras could reach that of conventional cameras, it could lead the way towards providing a cheaper automated DR diagnosis for diabetic people residing in (rural) Ethiopia.

\subsection{Performance of Deep Learning based Classifiers on Portable Fundus Cameras}

\citet{rogers2019evaluation} evaluated performance of a publicly available DL based DR classifier system, Pegasus \citep{visulytixlimited}, in identifying referable and proliferative DR, by performing binary classification on retinal images captured using a handheld fundus camera from 6404 patients. Although binary, \citet{rogers2019evaluation} also goes on to report that performance of the DL classifier on portable fundus cameras was comparable with its classification performance on images collected using conventional fundus cameras.

Although not publicly available, retinal images have been collected using portable fundus cameras for automated classifiers' performance evaluation \citep{rogers2019evaluation} and retinal images quality assessment purposes \citep{wang2015human}. To this date, and to the best of the author's knowledge, the CHASE\_DB1 \citep{chasedb1} is the only publicly available retinal pictures dataset, which have used handheld fundus cameras for data collection. The CHASE\_DB1 dataset contains only 28 retinal images, which is insufficient if DL algorithms are to be used for DR identification \citep{lim2019technical}.

\subsection{The Curse of Binary Classification}
\label{binary_classification}

In a study performed by \citet{warwick2017prevalence}, it has been seen that it can take more than 20 years until DR becomes sight threatening and requires medical treatment. Full DR diagnosis result that includes stage of the disease is necessary, in order for ophthalmologists to proceed with correct retinal treatment. Although the above offline mobile based DR diagnosis works can be seen as a good start, they don't provide severity scales (which can be one of normal, mild, moderate, severe, and proliferative according to \citet{wilkinson2003proposed}) of DR as diagnosis result, and they can't be used for early detection and treatment of the disease before vision loss or impairment is imminent.

\section{Challenge towards A Fully Offline Diabetic Retinopathy Diagnosis}
\label{proposed_methodology}

Quality of retinal pictures collected by fundus cameras' have been seen to be affected by Field of Views (FOV) and lens' diameter \citep{dehoog2009fundus}. Higher FOV and smaller diameter are preferred camera properties. Even though conventional fundus cameras that are currently on the market seem to achieve these specifications, which helps in classifying DR into different stages, portable fundus cameras output images of low quality; hence their use in binary classification only \citep{rogers2019evaluation, sosale2020medios}. Since identification of DR severity scale provides practitioners with the necessary information for treatment, automated fundus images classification systems should provide degree of DR by performing multi-class classification.

In the production of an offline, smartphone based, multi-class identifying DR diagnosis system for Ethiopia, the author recommends the development of a high quality portable fundus camera as the first priority over DL based fundus images classification. This is mainly because fundus images with reduced  quality were found to be unsuitable for automated and manual diagnosis of DR \citep{yao2016generic, lin2019retinal}. Producing high quality cameras would lead the way towards solving challenges associated with absence of medical equipments for diagnosing the disease \citep{foster2005impact}. It can be used as a mobile diagnostic equipment for manual DR grading. This can also help in collecting, and publicly providing dataset of fundus images, captured using handheld cameras, for the wider DL applications research community.

\section{Conclusion}
\label{conclusion}

The challenge associated with providing DR diagnosis in Ethiopia could be solved by combining high quality portable fundus cameras and mobile application development. Classification performance of automated DR diagnosis (into one of the five stages) has been seen to surpass manual grading \citep{gulshan2019performance}. After installing portable high quality fundus cameras to a smartphone, trained DR classifier models could be incorporated into a mobile application to provide instant diagnosis result. This approach could also be adapted to other underdeveloped countries with similar challenges.

\bibliography{iclr2020_conference}

\begin{thebibliography}{23}
\providecommand{\natexlab}[1]{#1}
\providecommand{\url}[1]{\texttt{#1}}
\expandafter\ifx\csname urlstyle\endcsname\relax
  \providecommand{\doi}[1]{doi: #1}\else
  \providecommand{\doi}{doi: \begingroup \urlstyle{rm}\Url}\fi

\bibitem[Abr{\`a}moff et~al.(2018)Abr{\`a}moff, Lavin, Birch, Shah, and
  Folk]{abramoff2018pivotal}
Michael~D Abr{\`a}moff, Philip~T Lavin, Michele Birch, Nilay Shah, and James~C
  Folk.
\newblock Pivotal trial of an autonomous ai-based diagnostic system for
  detection of diabetic retinopathy in primary care offices.
\newblock \emph{NPJ digital medicine}, 1\penalty0 (1):\penalty0 1--8, 2018.

\bibitem[DeHoog \& Schwiegerling(2009)DeHoog and
  Schwiegerling]{dehoog2009fundus}
Edward DeHoog and James Schwiegerling.
\newblock Fundus camera systems: a comparative analysis.
\newblock \emph{Applied optics}, 48\penalty0 (2):\penalty0 221--228, 2009.

\bibitem[Fite et~al.(2019)Fite, Lake, and Hanfore]{fite2019diabetic}
Robera~Olana Fite, Eyasu~Alem Lake, and Lolemo~Kelbiso Hanfore.
\newblock Diabetic retinopathy in ethiopia: A systematic review and
  meta-analysis.
\newblock \emph{Diabetes \& Metabolic Syndrome: Clinical Research \& Reviews},
  2019.

\bibitem[Foster \& Resnikoff(2005)Foster and Resnikoff]{foster2005impact}
Allen Foster and Serge Resnikoff.
\newblock The impact of vision 2020 on global blindness.
\newblock \emph{Eye}, 19\penalty0 (10):\penalty0 1133--1135, 2005.

\bibitem[Gosheva et~al.(2017)Gosheva, Klameth, Norrenberg, Clin, Dietter, Haq,
  Ivanov, Ziemssen, and Leitritz]{gosheva2017quality}
Mariya Gosheva, Christian Klameth, Lars Norrenberg, Lucien Clin, Johannes
  Dietter, Wadood Haq, Iliya~V Ivanov, Focke Ziemssen, and Martin~A Leitritz.
\newblock Quality and learning curve of handheld versus stand-alone
  non-mydriatic cameras.
\newblock \emph{Clinical ophthalmology (Auckland, NZ)}, 11:\penalty0 1601,
  2017.

\bibitem[Gulshan et~al.(2016)Gulshan, Peng, Coram, Stumpe, Wu, Narayanaswamy,
  Venugopalan, Widner, Madams, Cuadros, et~al.]{gulshan2016development}
Varun Gulshan, Lily Peng, Marc Coram, Martin~C Stumpe, Derek Wu, Arunachalam
  Narayanaswamy, Subhashini Venugopalan, Kasumi Widner, Tom Madams, Jorge
  Cuadros, et~al.
\newblock Development and validation of a deep learning algorithm for detection
  of diabetic retinopathy in retinal fundus photographs.
\newblock \emph{Jama}, 316\penalty0 (22):\penalty0 2402--2410, 2016.

\bibitem[Gulshan et~al.(2019)Gulshan, Rajan, Widner, Wu, Wubbels, Rhodes,
  Whitehouse, Coram, Corrado, Ramasamy, et~al.]{gulshan2019performance}
Varun Gulshan, Renu~P Rajan, Kasumi Widner, Derek Wu, Peter Wubbels, Tyler
  Rhodes, Kira Whitehouse, Marc Coram, Greg Corrado, Kim Ramasamy, et~al.
\newblock Performance of a deep-learning algorithm vs manual grading for
  detecting diabetic retinopathy in india.
\newblock \emph{JAMA ophthalmology}, 137\penalty0 (9):\penalty0 987--993, 2019.

\bibitem[Hagos et~al.(2019)Hagos, Kant, and Bala]{hagos2019automated}
Misgina~Tsighe Hagos, Shri Kant, and Surayya~Ado Bala.
\newblock Automated smartphone based system for diagnosis of diabetic
  retinopathy.
\newblock In \emph{2019 International Conference on Computing, Communication,
  and Intelligent Systems (ICCCIS)}, pp.\  256--261. IEEE, 2019.

\bibitem[IDF(2017)]{idf2015}
IDF.
\newblock Idf africa members, Apr 2017.
\newblock URL
  \url{https://www.idf.org/our-network/regions-members/africa/members/9-ethiopia.html}.

\bibitem[Kingston(2011)]{chasedb1}
University~Research Kingston.
\newblock Chase\_db1 database, Jan 2011.
\newblock URL \url{https://blogs.kingston.ac.uk/retinal/chasedb1/}.

\bibitem[Leasher et~al.(2016)Leasher, Bourne, Flaxman, Jonas, Keeffe, Naidoo,
  Pesudovs, Price, White, Wong, et~al.]{leasher2016global}
Janet~L Leasher, Rupert~RA Bourne, Seth~R Flaxman, Jost~B Jonas, Jill Keeffe,
  Kovin Naidoo, Konrad Pesudovs, Holly Price, Richard~A White, Tien~Y Wong,
  et~al.
\newblock Global estimates on the number of people blind or visually impaired
  by diabetic retinopathy: a meta-analysis from 1990 to 2010.
\newblock \emph{Diabetes care}, 39\penalty0 (9):\penalty0 1643--1649, 2016.

\bibitem[Lim et~al.(2019)Lim, Hsu, Lee, Ting, and Wong]{lim2019technical}
Gilbert Lim, Wynne Hsu, Mong~Li Lee, Daniel Shu~Wei Ting, and Tien~Yin Wong.
\newblock Technical and clinical challenges of ai in retinal image analysis.
\newblock In \emph{Computational Retinal Image Analysis}, pp.\  445--466.
  Elsevier, 2019.

\bibitem[Lin et~al.(2019)Lin, Yu, Weng, and Zheng]{lin2019retinal}
Jiawen Lin, Lun Yu, Qian Weng, and Xianghan Zheng.
\newblock Retinal image quality assessment for diabetic retinopathy screening:
  A survey.
\newblock \emph{Multimedia Tools and Applications}, pp.\  1--27, 2019.

\bibitem[oDocs Eye~Care(2018)]{oDocsnun2018}
oDocs Eye~Care.
\newblock odocs nun, Jul 2018.
\newblock URL \url{http://www.odocs-tech.com/odocs-nun/}.

\bibitem[Rajalakshmi et~al.(2018)Rajalakshmi, Subashini, Anjana, and
  Mohan]{rajalakshmi2018automated}
Ramachandran Rajalakshmi, Radhakrishnan Subashini, Ranjit~Mohan Anjana, and
  Viswanathan Mohan.
\newblock Automated diabetic retinopathy detection in smartphone-based fundus
  photography using artificial intelligence.
\newblock \emph{Eye}, 32\penalty0 (6):\penalty0 1138--1144, 2018.

\bibitem[Rogers et~al.(2019)Rogers, Gonzalez-Bueno, Franco, Star, Mar{\'\i}n,
  Vassallo, Lansingh, Trikha, and Jaccard]{rogers2019evaluation}
TW~Rogers, J~Gonzalez-Bueno, R~Garcia Franco, E~Lopez Star, D~M{\'e}ndez
  Mar{\'\i}n, J~Vassallo, VC~Lansingh, S~Trikha, and N~Jaccard.
\newblock Evaluation of an ai system for the detection of diabetic retinopathy
  from images captured with a handheld portable fundus camera: the mailor ai
  study.
\newblock \emph{arXiv preprint arXiv:1908.06399}, 2019.

\bibitem[Sosale et~al.(2020)Sosale, Sosale, Murthy, Sengupta, Naveenam,
  et~al.]{sosale2020medios}
Bhavana Sosale, Aravind~R Sosale, Hemanth Murthy, Sabyasachi Sengupta,
  Muralidhar Naveenam, et~al.
\newblock Medios--an offline, smartphone-based artificial intelligence
  algorithm for the diagnosis of diabetic retinopathy.
\newblock \emph{Indian Journal of Ophthalmology}, 68\penalty0 (2):\penalty0
  391, 2020.

\bibitem[Visulytix(2018)]{visulytixlimited}
Visulytix.
\newblock Pegasus, Nov 2018.
\newblock URL \url{https://www.f6s.com/visulytixlimited}.

\bibitem[Wang et~al.(2015)Wang, Jin, Lu, Cheng, Ye, and Qian]{wang2015human}
Shaoze Wang, Kai Jin, Haitong Lu, Chuming Cheng, Juan Ye, and Dahong Qian.
\newblock Human visual system-based fundus image quality assessment of portable
  fundus camera photographs.
\newblock \emph{IEEE transactions on medical imaging}, 35\penalty0
  (4):\penalty0 1046--1055, 2015.

\bibitem[Warwick et~al.(2017)Warwick, Brooks, Osmond, and
  Krishnan]{warwick2017prevalence}
Alasdair~N Warwick, Andrew~P Brooks, Clive Osmond, and Radhika Krishnan.
\newblock Prevalence of referable, sight-threatening retinopathy in type 1
  diabetes and its relationship to diabetes duration and systemic risk factors.
\newblock \emph{Eye}, 31\penalty0 (2):\penalty0 333--341, 2017.

\bibitem[WDF(2012)]{wdf2019}
WDF.
\newblock Establishing diabetes outpatient services wdf10-508, Dec 2012.
\newblock URL
  \url{https://www.worlddiabetesfoundation.org/projects/ethiopia-wdf10-508}.

\bibitem[Wilkinson et~al.(2003)Wilkinson, Ferris~III, Klein, Lee, Agardh,
  Davis, Dills, Kampik, Pararajasegaram, Verdaguer,
  et~al.]{wilkinson2003proposed}
CP~Wilkinson, Frederick~L Ferris~III, Ronald~E Klein, Paul~P Lee, Carl~David
  Agardh, Matthew Davis, Diana Dills, Anselm Kampik, R~Pararajasegaram, Juan~T
  Verdaguer, et~al.
\newblock Proposed international clinical diabetic retinopathy and diabetic
  macular edema disease severity scales.
\newblock \emph{Ophthalmology}, 110\penalty0 (9):\penalty0 1677--1682, 2003.

\bibitem[Yao et~al.(2016)Yao, Zhang, Xu, Fan, and Xu]{yao2016generic}
Zhenjie Yao, Zhipeng Zhang, Li-Qun Xu, Qingxia Fan, and Ling Xu.
\newblock Generic features for fundus image quality evaluation.
\newblock In \emph{2016 IEEE 18th International Conference on e-Health
  Networking, Applications and Services (Healthcom)}, pp.\  1--6. IEEE, 2016.

\end{thebibliography}
\bibliographystyle{iclr2020_conference}

\end{document}